\documentclass[aps,prb,a4paper,twocolumn,showpacs,showkeys,floatfix,superscriptaddress]{revtex4-1}

\usepackage[dvips]{epsfig}
\usepackage[dvips]{color}
\usepackage{amsmath}
\usepackage{multirow}

\newcommand{\figwidth}{0.98\columnwidth}
\newcommand{\vect}[1]{\mathbf{#1}}
\begin{document}
\title{ESR study of the spin ladder with uniform Dzyaloshinskii-Moria interaction.}

\author{V.N. Glazkov}
\email{glazkov@kapitza.ras.ru}

\affiliation{P.L.Kapitza Institute for Physical Problems RAS, Kosygin
str. 2, 119334 Moscow, Russia}

\affiliation{Moscow Institute of Physics and Technology, 141700
Dolgoprudny, Russia}

\author{M. Fayzullin}
\affiliation{Institute for Physics, Kazan Federal University, 18
Kremlyovskaya str. 18, 420008, Kazan, Russia}

\author{Yu. Krasnikova}
\affiliation{Kapitza Institute for Physical Problems RAS, Kosygin
str. 2, 119334 Moscow, Russia}

\affiliation{Moscow Institute of Physics and Technology, 141700
Dolgoprudny, Russia}

\author{G. Skoblin}
\altaffiliation{Current address: Quantum Device Physics
Laboratory, Department of Microtechnology and Nanoscience,
Chalmers University of Technology SE-412 96 G\"{o}teborg, Sweden }

\affiliation{Kapitza Institute for Physical Problems RAS, Kosygin
str. 2, 119334 Moscow, Russia}

\affiliation{Moscow Institute of Physics and Technology, 141700
Dolgoprudny, Russia}

\author{D.Schmidiger}
\affiliation{Neutron Scattering and Magnetism, Laboratory for
Solid State Physics, ETH Z\"{u}rich, 8006 Z\"{u}rich, Switzerland}

\author{S.M\"{u}hlbauer}
\altaffiliation{Current address: Heinz Maier-Leibnitz Zentrum
(MLZ) Garching, D-85748 Germany}
\affiliation{Neutron Scattering
and Magnetism, Laboratory for Solid State Physics, ETH Z\"{u}rich,
8006 Z\"{u}rich, Switzerland}

\author{A.Zheludev}
\affiliation{Neutron
Scattering and Magnetism, Laboratory for Solid State Physics, ETH
Z\"{u}rich, 8006 Z\"{u}rich, Switzerland}

\date{\today}

\begin{abstract}
Evolution of the ESR absorption in a strong-leg spin ladder magnet
(C$_7$H$_{10}$N$_2$)$_2$CuBr$_4$ (abbreviated as DIMPY) is studied
from 300~K to 400~mK. Temperature dependence of the ESR relaxation
follows a staircase of crossovers between different relaxation
regimes.  We ague that the main mechanism of ESR line broadening
in DIMPY is uniform Dzyaloshinskii-Moria interaction
($|\vect{D}|=0.20$~K) with an effective longitudinal component
along an exchange bond of Cu ions within the legs resulting from
the low crystal symmetry of DIMPY and nontrivial orbital ordering.
The same Dzyaloshinskii-Moriya interaction results in the lifting
of the triplet excitation degeneracy, revealed through the weak
splitting of the ESR absorption at low temperatures.

\end{abstract}

\keywords{low-dimensional magnet}

\pacs{75.10.Kt, 76.30.-v}

%75.10.Kt    Quantum spin liquids, valence bond phases and related phenomena
%76.30.-v Electron paramagnetic resonance and relaxation

\maketitle

\section{Introduction.}

Low-dimensional magnets are actively studied during last decades
both theoretically and experimentally. Spin-ladder is one of the
simplest models of the field, that is just one step more
complicated than the Heisenberg spin chain, the keystone of the
low-dimensional magnetism. Such a system consists of two chains
forming the "legs" of the spin ladder, which are coupled by simple
interchain coupling forming "rungs" of the ladder. The Hamiltonian
of the single spin ladder with the equivalent positions along the
ladder is
\begin{eqnarray}\label{eqn:ham}
{\cal H}&=&J_{leg}\sum_i (\vect{S}_{1,i}\vect{S}_{1,i+1}+\vect{S}_{2,i}\vect{S}_{2,i+1})+\nonumber\\
&&J_{rung}\sum_i \vect{S}_{1,i}\vect{S}_{2,i}+\mu_B \vect{H} \hat{g}\vect{S}+{\cal H}_{anis}
\end{eqnarray}
\noindent it includes Heisenberg exchange couplings $J_{leg}$ and
$J_{rung}$, Zeeman interaction (with usually anisotropic
$g$-tensor) and weak anisotropic spin-spin interactions ${\cal H}_{anis}$.

Independently on the ratio between the $J_{leg}$ and $J_{rung}$,
the excitation spectrum of the spin ladder is gapped, ground state
is non-magnetic and excited states are $S=1$
quasiparticles.\cite{kolezhukmikeska} However, most of the
experimentally available examples of the spin ladder systems are
the so-called strong-rung ladders with the dominating in-rung
interaction $J_{rung}$. Strong-leg ladders remains a rarity in
this family. Additional complication of the real systems is a
presence of the anisotropic spin-spin interactions breaking the
ideal symmetry of the Heisenberg model. Such interactions limit
excitations lifetime (to the point of total damping in some
extreme cases\cite{ntenp}) and could lift degeneracy of the $S=1$
states. Thus, estimation of such interactions strength and,
ideally, search for the systems with negligible anisotropic part
of the Hamiltonian is an important quest when comparing real
magnets with model predictions.

Adequate accounting for the effect of anisotropic interactions in
a spin-gap magnet is also a challenge. This problem was addressed
in a 1D field theory models via bosonic \cite{Affleck} and
fermionic \cite{tsvelik} approaches and in an independently
developed macroscopic model.\cite{farmar} However reliable
microscopic models remains a rarity (see, e.g. Ref.
\onlinecite{Kolezhuk}): most of the real spin-gap magnets have a
complicated network of the exchange couplings allowing far too
numerous possibilities of the anisotropic interactions parameters.
The adequate microscopic approaches are of particular interest in
connection with a particular case of the effect of a uniform
Dzyaloshinskii-Moria interaction on the properties of a quantum
magnet. \cite{povarov}

Recently found organometallic compound
(C$_7$H$_{10}$N$_2$)$_2$CuBr$_4$ , abbreviated DIMPY for short, is
an example of the strong-leg ladder with very weak anisotropic
interactions.\cite{dimpy1,dimpy2,dimpy-dave-prb,dimpy-dave-prl}
Presence of the energy gap in the excitation spectrum was revealed
by magnetic susceptibility \cite{dimpy1}, specific heat
\cite{dimpy2} and magnetization \cite{dimpy-dave-prl,dimpy-magn}
bulk measurements as well as by inelastic neutron scattering.
\cite{dimpy2,dimpy-dave-prb} The energy gap was found\cite{dimpy2}
to be 0.33 meV, it can be closed by the magnetic field $\mu_o
H_{c1}=3.0$~T, while the saturation field is much higher
\cite{dimpy-magn} $\mu_0 H_{sat} \approx 30$~T. The values of the
exchange constants were determined from the DMRG fit of the
measured inelastic neutron scattering spectra
\cite{dimpy-dave-prl} and were found to be $J_{leg}=1.42$~meV and
$J_{rung}=0.82$~meV. The magnetic field induced ordering is
observed at very low temperatures ($~T_N^{(max)} \approx 300$~mK
at $\mu_0 H \sim 15$~T).\cite{dimpy-dave-prl}

Electron spin resonance (ESR) spectroscopy is a powerful tool to
probe for the weak anisotropic interactions in the magnetic
systems. Inelastic neutron scattering experiments
\cite{dimpy-dave-prb} have shown that DIMPY is an almost perfect
realisation of the Heisenberg spin ladder. ESR technique allows
much higher energy resolution (routinely resolved ESR linewidth of
100 Oe corresponds approximately to the energy resolution of 1
$\mu$eV) and thus allows to probe possible effects of anisotropic
interactions with high accuracy.

In the present manuscript we report results of the ESR study of
low-energy spin dynamics in DIMPY in the temperature range from
400~mK to 300~K. We observe angular and temperature dependences of
the ESR linewidth at high temperatures which can be described as an
effect of the uniform Dzyaloshinskii-Moria (DM) interaction, which
is allowed by the lattice symmetry. At low temperatures we observe
slitting of the ESR absorption line due to lifting of the triplet
state degeneracy, which is also possibly due to the same DM
interaction. Additionally we observe well resolved ESR absorption
lines from the inequivalent ladders which allowed an upper estimate
of interladder exchange interaction.

\section{Samples and experimental details.}

Single crystals of non-deuterated DIMPY were grown from the solution
by slow diffusion in a temperature gradient. Samples quality was
checked by X-ray diffraction and magnetization measurements.
Concentration of the paramagnetic defects estimated from the 500~mK
magnetization curve is below 0.05\%.

DIMPY belongs to the monoclinic space group P2(1)/n with lattice
parameters $a=7.504$~\AA{}, $b=31.613$~\AA{}, $c=8.206$~\AA{} and
the angle $\beta=98.972^{\circ}$.\cite{dimpy1}  As-grown crystals
have a well developed plane orthogonal to the $b$-axis and are
elongated along the $a$ direction.

ESR experiments were performed using set of the home-made
transmission-type ESR spectrometers at the frequencies 18-38~GHz.
Lowest available temperature of 400~mK was obtained by He-3 vapours
pumping cryostat. At the measurements below 77~K magnetic field was
created by compact superconducting magnets, typical nonuniformity of
the magnetic field at the resonance conditions in our experiments is
estimated as $<5\ldots20$~Oe depending on the magnet used.
High-temperature experiments where done with a resistive
water-cooled magnet with the field nonuniformity about 5 Oe.

\section{Lattice symmetry and possible anisotropic interactions.\label{sect:sym}}

\begin{figure}
  \centering
  \epsfig{file=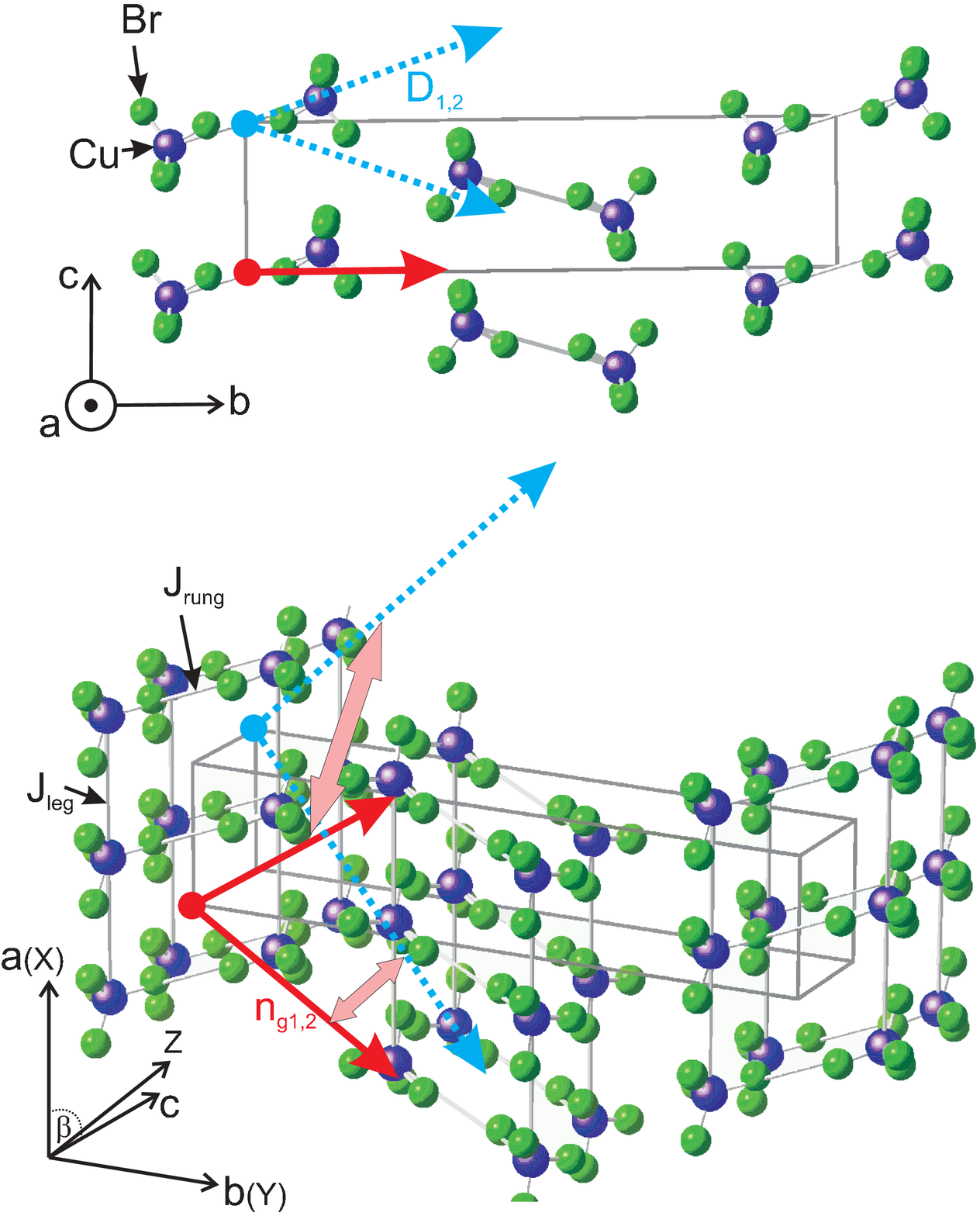, width=\figwidth, clip=}
  \caption{(color online) Crystallographic structure of DIMPY with two
  magnetically nonequivalent spin ladders. Only Cu and Br ions are
  shown along with the main exchange bonds $J_{leg}$ and
  $J_{rung}$. Solid arrows (red) indicate directions of $g$-tensor
  main axes for inequivalent ladders,
  dashed arrows (blue) indicate directions of the Dzyaloshinskii-Moriya
  vectors for inequivalent ladders, as found from the data fit (see
  text). Broad double-headed arrows links DM vector and $g$-tensor
  axis corresponding to the same ladder.}\label{fig:struct}
\end{figure}

Monoclinic unit cell of DIMPY includes four magnetic Cu$^{2+}$
ions that belongs to two spin ladders: two pairs of copper ions
form rungs of the spin-ladders, which are then reproduced by
translations along the $a$-axis. This results in the formation of
two ladders differently oriented with respect to the crystal
\cite{dimpy-dave-prb} (see Figure \ref{fig:struct}).

Space symmetry of the DIMPY lattice includes inversion center in
the middle of each rung and a second order screw axis parallel to
the crystallographic $b$ direction  that links different ladders.

These symmetries place strong restrictions on the possible
microscopic anisotropic interactions in DIMPY despite the low
crystallographic symmetry. First, all anisotropic interactions
along the legs of the ladders should be uniform because of
translational symmetry. Second, Dzyaloshinskii-Moria antisymmetric
interaction ${\cal H}_{DM}=\vect D \cdot \left[ {\vect S}_1 \times
{\vect S}_2 \right]$ on the rungs is forbidden by the inversion
symmetry. The same inversion symmetry requires that direction of
the Dzyaloshinskii-Moria vector $\vect D$ have to be exactly
opposite on the legs of the same ladder. Third, inversion center
on the rungs of the ladder ensures that $g$-tensor is always the
same for the given spin ladder so there are no complications of
anisotropic Zeeman splitting.

Second order axis establish relations between the $g$-tensor
components and Dzyaloshinskii-Moria vector direction in different
ladders. In particular, because of this second order axis the
effective $g$-factor values are the same for both ladders for the
field applied parallel or orthogonally  to this axis.

This analysis neglects anisotropic interladder couplings and
possible symmetric anisotropic exchange (SAE) coupling  ${\cal
H}_{SAE} = \sum _{\mu,\tau} J_{\mu\tau} { S}_1^\mu {S}_2^\tau$,
where $J_{\mu\tau}$ are components of a symmetric exchange tensor
$\hat{\mathbf{A}}$, which is usually constrained by condition $Tr
J_{\mu,\tau}=0$. Symmetric interaction is allowed both on rungs and
legs of the ladder and is also constrained by symmetry operations.
However, as we will demonstrate below, our observations point that
Dzyaloshinskii-Moria interaction is dominating anisotropic
interaction in the case of DIMPY.

As for the anisotropic interladder couplings, there is a
possibility that anisotropic couplings between the equivalent
ladders stacked in $c$ direction could be important as well: Cu-Cu
distance in this direction is even less then the distance on the
rungs ($8.2$~\AA{} against $8.9$~\AA{}) and suppression of the
Heisenberg exchange interaction in this direction is most likely
due to unfavorable mutual orientation of the electron orbitals of
bromine ions mediating this superexchange route which could be
less important for the anisotropic spin-spin interactions arising
through involvement of differently oriented excited electron
orbitals mixed with the ground state by spin-orbital
interaction.\cite{eremins} In the present work we neglect this
possibility.

Thus, the main anisotropic interactions in DIMPY are really simple
to analyse. They include anisotropic $g$-tensor, which is the same
for all magnetic ions of the given spin-ladder and a
Dzyaloshinskii-Moria interaction, which is uniform along the leg
of the ladder and the Dzyaloshinskii-Moria vectors are exactly
opposite on the legs of the given ladder.

\section{Experimental results.}
\subsection{Angular dependence of the ESR absorption at 77~K.}
\begin{figure}
  \centering
  \epsfig{file=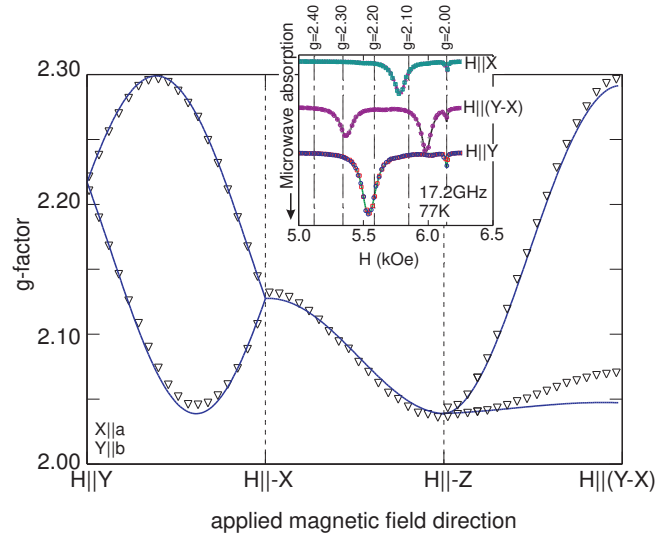, width=\figwidth, clip=}
  \caption{(color online) Main panel: angular dependence of the $g$-factor at
  77~K. Symbols - experimental data, curves - uniaxial $g$-tensor
  best fit (see text). Experimental error is about of 0.1\% and is
  within symbol size. Inset: example of ESR absorption spectra at
  representative orientations. Symbols - experimental data, curves
  - best fit with Lorentzian lineshape, vertical lines corresponds
  to certain $g$-factor values. Narrow line with $g=2.00$ (at 6.13
  kOe) is a DPPH marker.    }\label{fig:g-angular}
\end{figure}

\begin{figure}
  \centering
  \epsfig{file=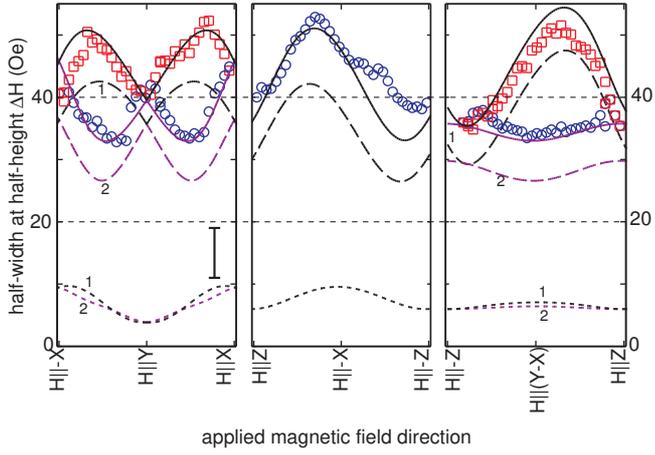, width=\figwidth, clip=}
  \caption{(color online) Angular dependence of the ESR linewidth at f=17.2~GHz,
  T=77~K (half-width at half-height). Vertical bar at the left panel shows typical
  errorbar size (double error). Squares: width of the high-$g$ component.
  Circles: width of the low-$g$ component. Curves: model
  description (see text), solid lines show full linewidth, dashed
  and dotted lines show contributions due to DM and SAE
  interactions respectively. Marks "1" and "2" indicate
  contributions corresponding to the same
  ladder.}\label{fig:width-angular}
\end{figure}

We have taken rotational patterns of ESR absorption for the
magnetic field applied in different crystallographic planes.
Because of monoclinic lattice symmetry care should be taken with
consistent determination of the field direction with respect to
the lattice axes. We will use a cartesian basis with $X||a$,
$Y||b$ and $Z||c^{{}*{}}$ for the direction description. Rotation
patterns were taken for the field confined to $(XY)$ and $(XZ)$
planes and to the plane containing $Z$ axis and an $(Y-X)$
direction. All rotation patterns were taken for more then $180^0$
angular sweeps.

Examples of absorption spectra and angular dependences of the
$g$-factor are shown of the Figure \ref{fig:g-angular}. We observe
one or two Lorentzian absorption lines. These absorption lines are
clearly due to the different spin ladders: the ladders are
equivalent with respect to the magnetic field for $\vect{H}||Y$
and $\vect{H} \perp Y$, and we observe single component absorption
at these orientations. Anisotropy of $g$-factor is typical for the
Cu$^{2+}$ ion, $g$-factor varies from about 2.03 to 2.30 in
agreements with powder ESR measurements of
Ref.\onlinecite{dimpy-highfieldESR}

Angular dependence of the ESR linewidth was determined by fitting
observed absorption spectra with a single lorentzian line or with
a sum of two lorentzian lines (Fig.\ref{fig:width-angular}).
Typical half-width at half-height at 77~K was around 50 Oe. Field
inhomogeneity in the used magnet and uncertainties of the fit
procedure limit accuracy of the linewidth determination to about
5~Oe, however angular dependence is clearly present. We were able
to cross-check our results at certain selected orientations on a
commercial Bruker X-band spectrometer and we have found that
X-band data are in agreement with our results.

\subsection{Low-temperature ESR.}
\begin{figure}
  \centering
  \epsfig{file=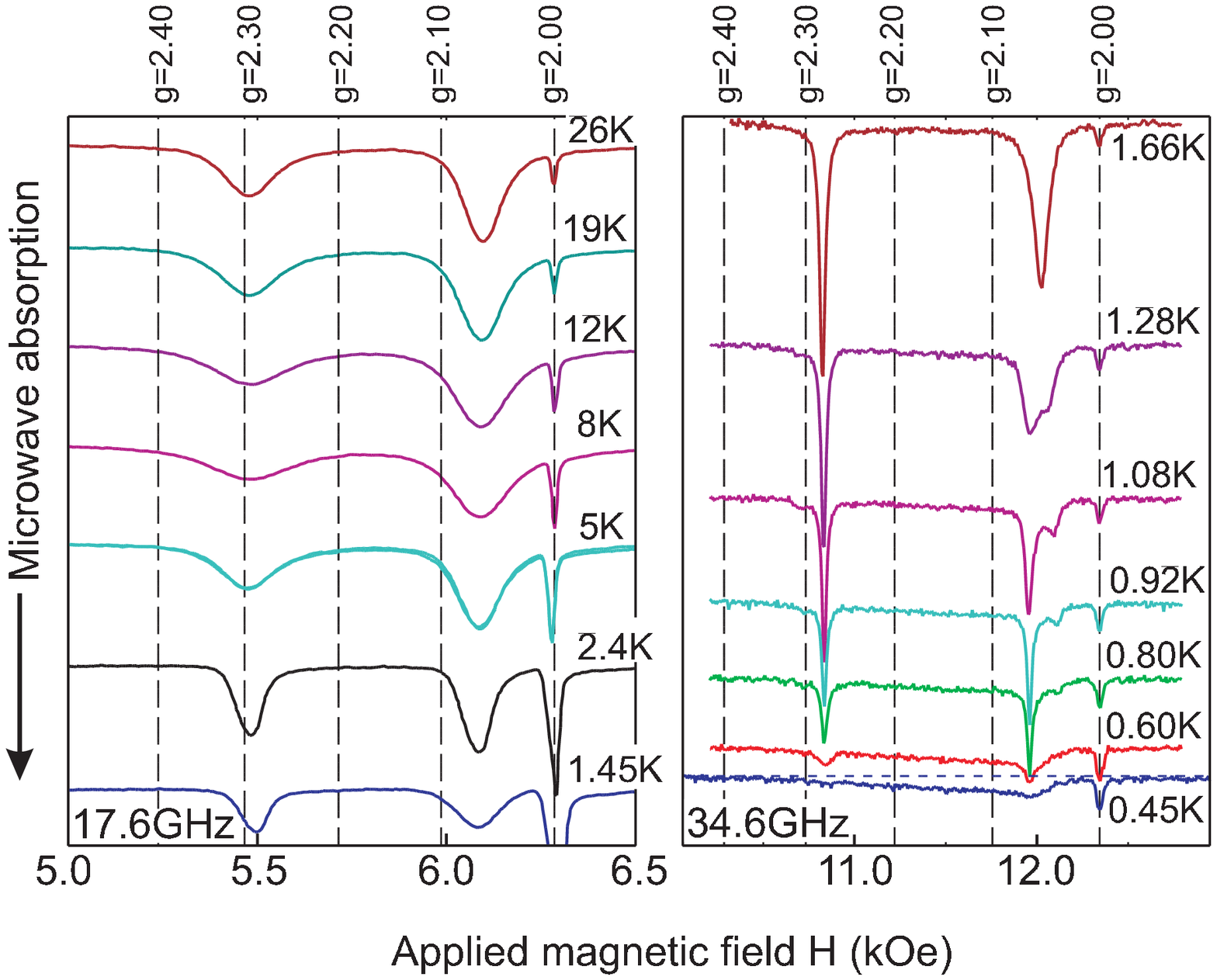, width=\figwidth, clip=}
  \caption{(color online) ESR absorption spectra at low
  temperatures, $\vect{H}||(X+Y)$. Vertical
  dashed lines mark resonance fields corresponding to the shown
  $g$-factor values. Horizontal dashed line at 0.45~K curve is a
  guide to the eye at zero-absorption level. Narrow absorption line at $g=2.00$ is a DPPH marker.}\label{fig:scans-lt}
\end{figure}

\begin{figure}
  \centering
  \epsfig{file=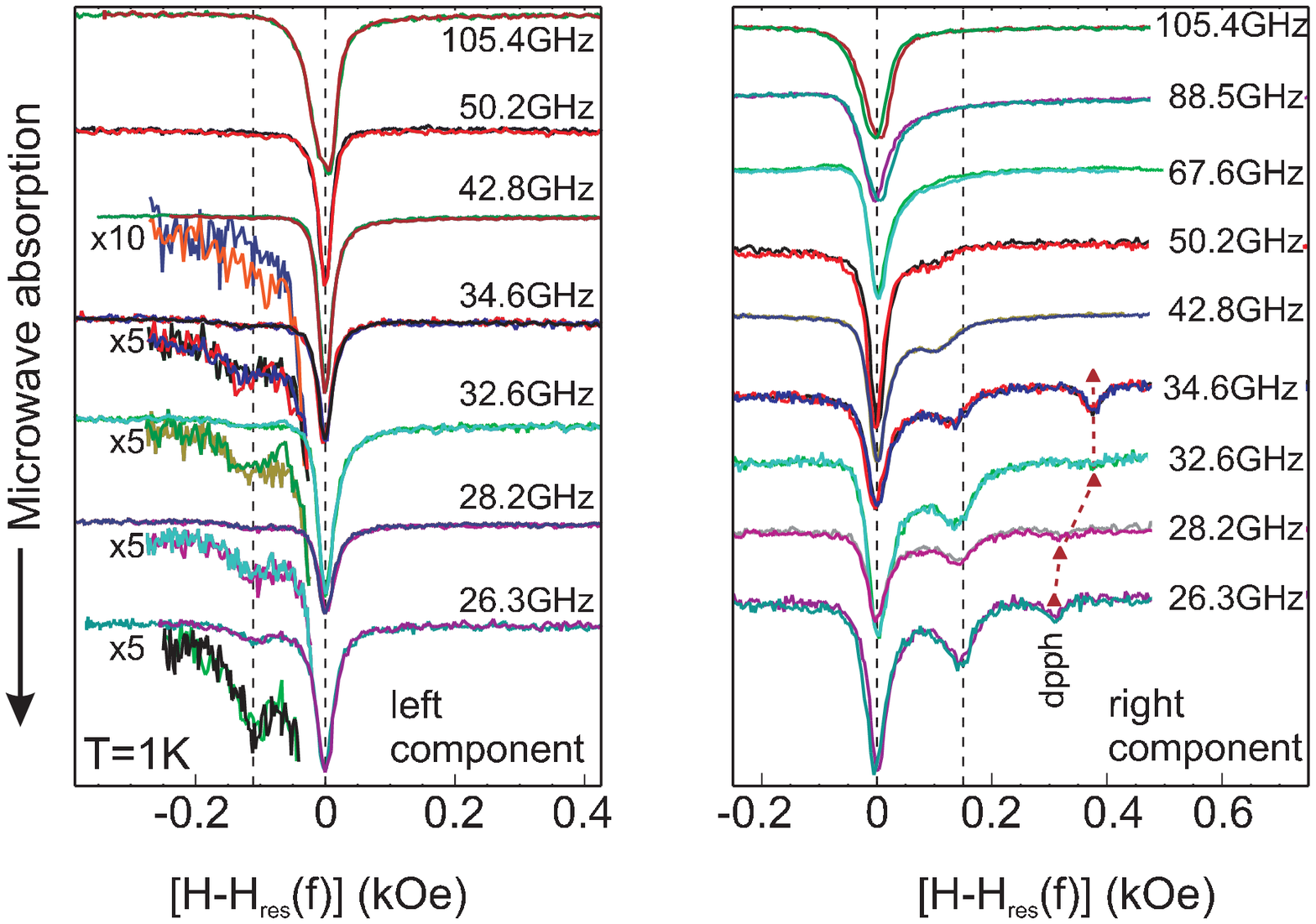, width=\figwidth, clip=}
  \caption{(color online) ESR absorption spectra at the temperature $~T \approx 1$~K
  at different frequencies, $\vect{H}||(X+Y)$.  All spectra are shifted along the field
  axis to fit positions of the main absorption subcomponents. Left
  panel: left absorption component ($g \approx 2.28$), weak
  absorption subcomponent is magnified by the factor of 5 or 10
  for better presentation. Right panel:
  right absorption component ($g \approx 2.05$). Vertical dashed
  lines mark positions of the absorption subcomponent at lowest
  frequency. Triangles on the right panel mark position of the
  DPPH marker absorption ($g=2.00$).  }\label{fig: max-split}
\end{figure}

\begin{figure}
  \centering
  \epsfig{file=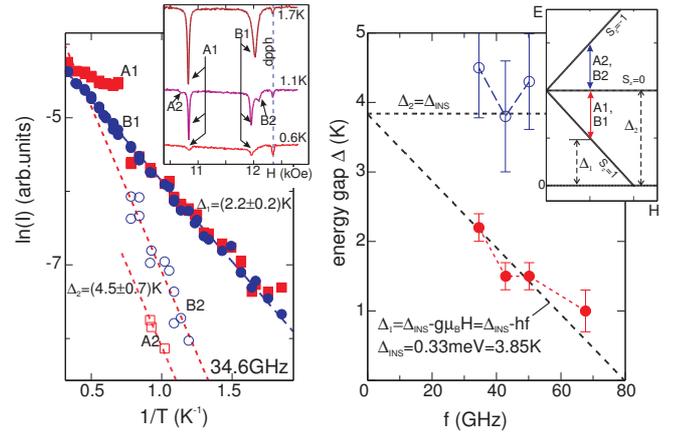, width=\figwidth, clip=}
  \caption{(color online) Left panel: temperature dependence of the ESR intensity
  below 1~K at f=34.6~GHz, $\vect{H}||(X+Y)$. Inset: examples of ESR absorption and ESR
  components and subcomponents notations. Symbols - experimental
  data, dashed lines - fits with thermoactivation law
  $I\propto \exp(-\Delta/T)$. Right panel: dependence of the determined
  activation gaps for different spectral subcomponents. Filled
  symbols: intense A1 and B1 subcomponents, open symbols - weak
  A2 and B2 subcomponents. Lines: parameters-free model
  dependence calculated with the zero-field gap value
  $\Delta_{INS}$ known from the inelastic neutron scattering
  experiments.\cite{dimpy2,dimpy-dave-prb}
  Inset: scheme of the energy levels of a spin-gap magnet in a
  magnetic field. Solid vertical arrows show transitions
  corresponding to the observed ESR absorption, dashed vertical
  arrows mark activation gaps for these transitions.
  }\label{fig:int-final}
\end{figure}

Low temperature (below 77~K) ESR absorption was measured at certain
fixed field directions: for the field applied parallel to the
symmetry axis $\vect H||Y$ and for the field canted by approximately
$45^0$ towards $X$ axis. In the first case both ladders are
equivalently oriented with respect to the magnetic field, while the
later case corresponds to the maximal difference of the ladders'
effective $g$-factor, as evidenced by 77~K measurements.

As expected, we observe single-component ESR absorption for $\vect
{H}||Y$ and two resolved ESR signals from different ladders for the
canted sample. Temperature evolution of the ESR absorption spectra
is qualitatively similar in both cases (Figure \ref{fig:scans-lt}).
Below 10~K the ESR absorption intensity freeze down due to the
presence of the energy gap. ESR signal continue to loose intensity
down to 450~mK and almost vanishes at this temperature. Lowest
temperature (450 mK) ESR absorption includes broad powder-like
absorption spectrum probably related to the distorted surface of the
sample.

We did not observed any additional absorption signals which could be
related to the formation of the field induced ordered phase above
the critical field to appear at the lowest temperature of 450 mK in
the fields up to 10~T at the frequencies of 26...35~GHz. This is in
agreement with the known phase diagram of DIMPY
\cite{dimpy-dave-prl} demonstrating that highest temperature of the
transition into the ordered state is about 300 mK.

Additional splitting of the ESR absorption lines was observed around
1~K (Figure \ref{fig: max-split}), resonance fields of the split
sub-components differ by approximately 150 Oe. This splitting was
observed at various frequencies, it was most pronounced on the
high-field component of the canted sample ESR absorption spectra.
One of the split sub-components is much weaker then the other and
freeze out faster on cooling. Remarkably, mutual orientation of the
weaker and stronger sub-components is different for the low-field
and high-field components. We did not observe resolved splitting for
the $\vect{ H}||Y$ orientation, instead a weak peak of the linewidth
was observed around the same temperature of 1~K probably indicating
unresolved splitting.

At low temperatures intensities of all components follow
exponential law $I\propto\exp(-\Delta/T)$ (Figure
\ref{fig:int-final}). Energy gap for the weaker sub-components is
larger then that for the main subcomponents. By taking temperature
dependences of the ESR absorption at different frequencies we were
able to determine the values of the energy gaps at several
frequencies revealing dependence of the activation energy from the
resonance field (Figure \ref{fig:int-final}).

\subsection{ESR linewidth evolution from 300~K to 400~mK.}
\begin{figure}
  \centering
  \epsfig{file=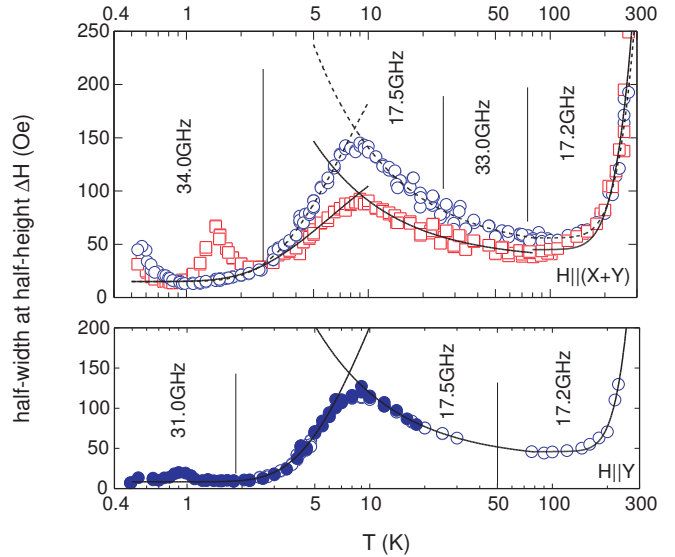, width=\figwidth, clip=}
  \caption{(color online) Temperature dependence of the ESR
  linewidth. Upper panel: $\vect{H}||(X+Y)$. Circles: low-field
  (high-$g$) component, squares: high field (low-$g$) component.
  All data are measured on the samples from the same batch.
  Lower panel: $\vect{H}||Y$. Filled and open symbols corresponds to the data
  measured on samples from different batches.
  Experimental data were collected in different experimental setups operating at
  different frequencies, vertical lines mark approximate temperature
  boundaries for different experiments, microwave frequencies of
  each experiment are given. Typical errorbar size is around
  symbol size. Curves on both panels show empirical fit equations
  (see text and Table \ref{tab:fits}). }\label{fig:width_t}
\end{figure}

Temperature evolution of the ESR linewidth was measured from room
temperature down to 400~mK (Figure \ref{fig:width_t}). The
temperature dependence is qualitatively similar in all
orientations and demonstrate strongly non-monotonous behavior. At
high temperatures (above 90~K) linewidth strongly increases with
heating rising from about 50 Oe at 77~K to about 300 Oe at 300~K.
On cooling below 77~K linewidth again increases reaching maximum
at temperature $~T_{max}=9.0\pm 0.2$~K. Temperature of the maximum
is the same for all orientations, while linewidth value at the
maximum varies from 90~Oe to 140~Oe, both of the extreme values
being observed in the orientation of maximal splitting
$\vect{H}||(X\pm Y)$ for different components of ESR absorption.
Below $~T_{max}$ linewidth again decreases reaching its minimal
value of about 10~Oe (observed at $\vect H||Y$) at 2~K, which is
most likely limited by the field inhomogeneity in our setup. On
cooling below 2~K a peak in the linewidth is observed around 1~K.
The peak is most pronounced for the high-field component in the
orientation of maximal splitting of the ESR absorption components,
peak position coincides with the temperature of subcomponents
appearance. Similar but less pronounced peak is observed for
$\vect{H}||Y$. High-$g$ component in the $\vect{H}||(X+Y)$
orientation does not demonstrate such a peak, which is probably
related to the very low intensity of the appearing weaker
subcomponent which cause fitting procedure to lock on the main
spectral subcomponent. Finally, on cooling below 1~K linewidth of
both components in the $\vect{H}||(X+Y)$ orientation increases
again.

\section{Discussion.}
\subsection{Recovery of the $g$-tensor.\label{sec:g-tensor}}
Observed angular dependences of the $g$-factor can be fitted
assuming uniaxial $g$-tensor. As was described in the Section
\ref{sect:sym} $g$-tensor is the same for the given ladder and
orientations of the $g$-tensors in inequivalent ladders are bound
by 2-nd order axis. Hence, directions of the main axis can be
expressed via polar angles as $\vect {n}_{g1,2}=(\pm \sin \Theta
\cos\phi; \cos\Theta;\pm \sin\Theta \sin\phi)$, here we count
polar angle $\Theta$ from the 2-nd order axis $Y||b$, different
signs corresponds to the different ladders.

Least squares fit of our data (see Figure \ref{fig:g-angular})
yields $g$-tensor components $g_{\parallel}=2.296\pm0.010$ and
$g_{\perp}=2.040\pm0.006$ and angles $\Theta=(34.8\pm 1.5)^\circ$
and $\phi=(178\pm 4)^\circ$. Fit quality can be improved by
assuming general form of the $g$-tensor. However, this results
only in  minor planar anisotropy (with principal $g$-factor values
of $2.038\pm 0.010$ and $2.058\pm 0.010$) which is on the edge of
experimental error.

Main axis of the $g$-tensor within accuracy of our experiment lies
in the $(XY)$-plane of the crystal. This seems to be accidental as
there is no symmetry reasons to choose this plane in the
monoclinic crystal. Orientation of the $g$-tensor main axes
$\vect{n}_{g1,2}$ with respect to the crystal structure is shown
at the Figure \ref{fig:struct}. We can not decipher which of the
orientations corresponds to different ladders.

Found values of the main $g$-tensor components coincide with the
values found in earlier powder high-field ESR experiment.
\cite{dimpy-highfieldESR} The value of the $g$-factor for the
$\vect{H}||a$ case $g_a=2.17$ was found in
Ref.\onlinecite{dimpy-dave-prl} by magnetization fit, this value
disagree by 2\% with the value $g_a^{(ESR)}=2.130\pm0.005$ found
in our experiment.

\subsection{Interladder coupling estimation.}
Anisotropy of the $g$-tensor opens a direct way to the estimation
of interladder coupling. If the coupling between the ladders with
different g-tensor orientation would be strong enough, then a
common spin precession mode would be observed, a well known
exchange narrowing phenomenon.\cite{anderson-stat,exnar2}

Instead, we observe well resolved ESR absorption signals from the
inequivalent ladders. Thus, upper limit on the interladder coupling
can be estimated from the minimal splitting $\Delta H$ observed,
which is around 40 Oe at 17~GHz experiment (corresponds to
components $g$-factor difference $\Delta g=0.015$ at Figure
\ref{fig:g-angular}).

As it is known from the exchange narrowing theory, equally intense
components with splitting $\Delta\omega$ will form a common
precession mode if the coupling strength is above $J_c \simeq
\hbar\Delta\omega$. Hence we obtain an estimate $J_{inter}<{h\nu}
\frac{\Delta H}{H_{res}}\simeq 5$~mK or about 0.5 $\mu$eV.

This value is in reasonable agreement with the earlier estimate of
the interladder exchange coupling from the ordered phase boundary
calculated in mean-field approximation\cite{dimpy-dave-prl} as $n
J'_{MF}=6.3~\mu$eV (here $n$ is the number of coupled ladders,
coupling considered to be equal in all directions). Note that our
observation provides a direct estimate of the coupling between the
unequivalently oriented ladders only.

\subsection{Linewidth temperature dependence.\label{sec:width-t}}

\begin{table*}
\caption{ESR linewidth empirical fit equations parameters in
different temperature ranges and at different orientations. For
the orientations with two ESR components resolved (LF) and (HF)
marks fit results for the low-field and high-field components
correspondingly, (HF+LF) marks cases where linewidths of both
components are close and their temperature dependences are fitted
jointly. \label{tab:fits}}
\begin {ruledtabular}
\begin{tabular}{lccccc}
    {Temperature range}&Parameters & Typical fit &$\vect {H}||Y$&$\vect{H}||(x+ y)$&$\vect{H}||(y + z)$\\
    {and fit eqns.} & &accuracy & & &\\
    \hline
    $~T>80$~K& $\Delta H_0$, Oe& $\pm 5$~Oe  & 46& 45 (HF); 56 (LF) & \\
    $\Delta H=\Delta H_0+A \exp(-E_a/T)$& $A$,  $\times 10^4$~Oe&$\pm 50$\% & $5.9 $&$1.7 $ (HF); $1.8$ (LF)  & \\
    & $E_a$, K&$\pm 150$~K &1510 & 1240 (HF); 1310 (LF)& \\
    \hline
    15K$<T<80$~K& $\Delta H_\infty$, Oe& $\pm 5$~Oe &35 &35 (HF); 44(LF) & 40 (HF+LF)\\
    $\Delta H=\Delta H_\infty(1+\Theta/T)$&$\Theta$, K&$\pm 3$~K & 24& 16 (HF); 22(LF) & 14 (HF+LF)\\
    \hline
    1.5K$<T<7$~K&$\Delta H_0$, Oe&$\pm 2$~Oe & 10 &15 (HF);15 (LF)& 15(HF+LF)\\
    $\Delta H=\Delta H_0+A \exp(-E_a/T)$& $A$,  $\times 10^2$~Oe&$\pm 30$\% & 6.4 &1.7 (HF);3.9 (LF) & 5.3 (HF+LF)\\
    &$E_a$, K&$\pm 1.5$ & 14 (9.6\footnote{Value of $E_a=9.6~K$ corresponds to the best fit with fixed parameter $\Delta H_0=0$ (see text).}) & 6.4 (HF);8.4 (LF) & 10.5(HF+LF)\\

\end{tabular}
\end{ruledtabular}
\end{table*}

Non-monotonous temperature dependence
of the linewidth indicates that spin precession relaxation is
governed by different processes in the different parts of the
studied temperature range. We fit this temperature dependence by
set of empirical equations as shown on Figure \ref{fig:width_t}
and discussed below. Values of the fit coefficients for the
empiric equations used are gathered in the Table \ref{tab:fits}.

High temperature increase of the ESR linewidth (above 77~K) is
naturally related to the spin-lattice relaxation: increase of the
phonons population numbers leads to the increase of the relaxation
rate. Linewidth dependences can be fitted by the sum of the
constant contribution describing the high-temperature spin-spin
relaxation and an empirical activation law $\Delta H=\Delta H_0+A
\exp(-E_a/T)$. Activation energy is $E_a=(1400\pm150)K$. Similar
behavior with activation energy of the same order of magnitude was
reported for other cuprates \cite{marsel,monika} and it was
discussed \cite{monika} as a relaxation via excited state with a
competing Jan-Teller distortion. However, detailed analysis of the
lattice relaxation is beyond the scopes of the present paper.

Lattice contribution vanishes with cooling. Shallow minimum of the
linewidth at 70-100~K indicates that phonon relaxation channel is
practically frozen down here. Hence, we assume that linewidth
measured at 77~K is mostly due to spin-spin relaxation.

It is well known \cite{Abragam, AltKoz} that anisotropic spin-spin
interactions are responsible for the spin-spin relaxation. Thus,
ESR linewidth provides access to determine these interactions
strength. For the concentrated magnets Dzyaloshinskii-Moria
interaction (which is allowed by the lattice symmetry of DIMPY)
and symmetric anisotropic exchange interaction are the main
contributions. Temperature dependence of the ESR linewidth is one
of the physical effects to find which of theses anisotropic
interactions dominates the linewidth.

Detailed description of the ESR linewidth in a quantum spin-ladder
is only emerging now: case of spin ladder with symmetric
anisotropic spin-spin coupling was considered by recently by
Furuya and Sato \cite{Sato}, a theory accounting for the uniform
Dzyaloshinskii-Moriya interaction is still to be constructed.
Theory of an ESR linewidth for a quantum $S=1/2$ chain was
developed by Oshikawa and Affleck more then decade
ago.\cite{oshikawa} We will apply their results to understand
qualitatively temperature dependence of the ESR linewidth at high
temperatures $~T\gg J_{leg,rung}$. Oshikawa and Affleck have
demonstrated  that contribution of the symmetric anisotropic
exchange interaction to the ESR linewidth (exchange anisotropy in
their terms) decreases with cooling. They also considered
contribution of the staggered Dzyaloshinskii-Moria interaction and
have found that in this case ESR linewidth is increasing as
$1/T^2$ at low temperatures. Oshikawa and Affleck demonstrated
that at high temperature limit staggered Dzyaloshinskii-Moria
interaction results in the linewidth increasing with cooling as
$1/T$. As the high-temperature linewidth is determined by the pair
spin correlations, this result should be actually the same for the
staggered and uniform Dzyaloshinskii-Moria interaction. This
conclusion is in agreement with the results of
Ref.\onlinecite{marsel} for the uniform Dzyaloshinskii-Moria
interaction in quasi one dimensional antiferromagnet
Cs$_2$CuCl$_4$. Thus, increase of the linewidth with cooling below
80~K is a direct indication of the dominating role of the
Dzyaloshinskii-Moria interaction for the spin relaxation processes
in DIMPY. To model first order of the $1/T$ expansion we fit our
data by the law $\Delta H=\Delta H_\infty (1+ \Theta/T)$.
Characteristic temperature $\Theta$ is anisotropic and varies from
15 to 25~K in the orientations presented on the Figure
\ref{fig:width_t}. This temperature scale is close to the
exchange integral value in agreement with the results of
Refs.\onlinecite{oshikawa},\onlinecite{marsel}.

The crude estimate of this interaction strength can be obtained
from the linewidth at 80~K, which is approximately 50 Oe. As this
temperature far exceeds the exchange integral scale
high-temperature approximation can be used. We will discuss exact
calculations below while describing angular dependence, but as an
estimate one can write $\hbar \Delta\omega \sim \frac {D^2}{J}$ or
$D \sim \sqrt{g\mu_B \Delta H J} \sim 0.3$~K.

As the temperature decreases below approximately 10~K linewidth
start to decrease. This decrease is naturally related to the
gapped spectrum of the spin ladder. At low temperatures magnetic
properties of a spin ladder can be described on the triplet
quasiparticles language and linewidth is then interpreted as an
inverse lifetime of these quasiparticles, which is partially
determined by their interaction. As temperature approaches scale
of the energy gap, quasiparticles population numbers decreases,
gas of the quasiparticles became diluted and quasiparticles
interaction contribution froze out. This results in the narrowing
of the ESR absorption line with cooling. The linewidth temperature
dependence indeed follows thermoactivation law $\Delta H =\Delta
H_0+A \exp(-E_a'/T)$ with activation energy $E_a'=6.4...14$~K in
different orientations.

If the relaxation processes would be due to the pair interaction
of quasiparticles, the relaxation rate would be proportional to
the quasiparticles concentration squared and the activation energy
would be about $2\Delta_0$, where $\Delta_0=0.33$ meV (or 3.8~K)
is a zero-field gap. The activation energies for $\vect{H}||(X+Y)$
orientation are close to this expectation. However, for the field
applied along the second-order axis $\vect{H}||Y$ the best fit
activation energy ($14\pm1.5$~K) far exceeds the doubled
zero-field gap  value. Most likely this result is an artefact due
to the effects of field inhomogeneity in our experimental setup:
the low temperature linewidth is minimal for $\vect{H}||Y$ and
could be limited by experimental resolution. This leads to
overestimation of $\Delta H_0$ parameter which in turn results in
overestimation of the activation energy. Tentative fit of the
$\vect{H}||Y$ data with $\Delta H_0$ value fixed to zero yields
activation energy of 9.6~K which is much closer to twice
zero-field gap value. Similar $\Delta H_0=0$ fits in other
orientations lead to the smaller corrections of the determined
activation energy.

Thus, within accuracy of our experiment, which is mostly limited
by field inhomogeneity of the magnetic field in the
superconducting coil used, we can conclude that pair
quasiparticles interactions dominates spin-spin relaxation
processes of the spin-ladder in low-temperature regime.

The peak of the linewidth around 1~K is related to the splitting of
the ESR lines into subcomponent, its origin is related to the
classical exchange narrowing phenomenon.\cite{anderson-stat,exnar2}
The exchange frequency became temperature dependent being related to
the quasiparticles concentration. At low temperatures (low
quasiparticles concentration) split ESR line is observed, at higher
temperatures (higher quasiparticles concentration) effective
exchange interaction between the quasiparticles gain efficiency and
a common precession mode is formed. Crossover between these regimes
results in the broadening of ESR line. Similar effect is observed in
other spin-gap magnets \cite{glazkov-tlcucl3,glazkov-phcc} and in
other systems. \cite{chestnut}

Finally, definitive increase of the linewidth below 700~mK is
probably indicative of the critical regime in the vicinity of the
field induced phase transition.

\subsection{Angular dependence of the ESR linewidth.}\label{sec:width-ang}

According to the theory of the exchange narrowed resonance spectra
\cite{Anderson1953,Castner1971}, the half width at half maximum
for a single Lorentzian shaped line is given by
\begin{equation}
\Delta H = C\left[\frac{M_{2}^{3}}{M_{4}}\right]^{1/2},
\label{linewidth}
\end{equation}
\noindent where $C$ is dimensionless constant of order unity,
depending on how the wings of Lorentzian profile drop at fields of
the order of exchange field ($J/g\mu_{\rm B} \ll \Delta
H$)\cite{Castner1971}; $M_2$ and $M_4$ are the second and fourth
moments of resonance line, firstly introduced by Van Vleck
\cite{VanVleck1948}
\begin{equation}
M_2 =\frac{\langle [{\cal H}_{anis},S^+][S^-,{\cal H}_{anis}]\rangle}{h^2\langle S^+S^-\rangle}, \label{M2}
\end{equation}
\begin{equation}
M_4 = \frac{\langle [{\cal H}_{ex},[{\cal H}_{anis},S^+]][[S^-,{\cal H}_{anis}],{\cal H}_{ex}] \rangle}{h^4\langle S^+S^-\rangle}. \label{M4}
\end{equation}
where $S^{\pm}$ denote left/right circular components of the total
spin summed up over the whole sample, ${\cal H}_{ex}$ is isotropic
exchange Hamiltonian, ${\cal H}_{anis}$ is anisotropic one that
doesn't commute with ${\cal H}_{ex}$ hence causing broadening of the
resonance line.

Analysis of the ESR linewidth based on calculation of the spectral
moments is a well developed method which allows to identify nature
of spin-spin interactions and estimate their magnitudes in
magnetically concentrated systems.\cite{Zakharov2008} Its benefit
is that in high temperature limit ($T\to\infty$) an exact
expression for linewidth can be found out for an arbitrary spin
system, whatever space dimension and exchange
couplings.\cite{AltKoz,Huber1999}

In the present paper we apply the "method of moments" to a
strong-leg spin ladder system, described by Hamiltonian
(\ref{eqn:ham}) with uniform Dzyaloshinskii-Moriya interaction
\begin{equation}
{\cal H}_{DM}=\sum_i \sum_{l=1,2}\vect{D}_l
[\vect{S}_{l,i}\times\vect{S}_{l,i+1}],\label{eqn:Hanis}
\end{equation}

\noindent here $i$ enumerates rungs of the ladder and $l$ enumerates
legs of the ladder, DM vectors on the legs are considered arbitrary
for the moment ($\vect{D}_1\neq\vect{D}_2$). Substituting Eq.
(\ref{eqn:Hanis}) into Eqs.(\ref{M2}), (\ref{M4}) and using the
corresponding commutation relations for $S=1/2$ spin operators, for
the linewidth Eq. (\ref{linewidth}) in high temperature limit we
have
\begin{equation}
\Delta H_{\infty}^{DM}(Oe)=C\sum_{l=1,2}\frac{[D_x^2+D_y^2+2D_z^2]_l}{4\sqrt{2}\mu_{B}\tilde{J}_Dg(\theta,\phi)}
\label{eqn:dH}
\end{equation}

\noindent where $\tilde{J}_D=\sqrt{J_{leg}^2+2J_{rung}^2}$ mean an
average exchange integral, \cite{Huber1999} and angular dependence
is determined by the transformation for DM vector
\begin{eqnarray}
D_x &=& D_X\cos{\beta}\cos{\alpha}+D_Y\cos{\beta}\sin{\alpha}-D_Z\sin{\beta},\nonumber \\
D_y &=& D_Y\cos{\alpha}-D_X\sin{\alpha}, \label{Dtr} \\
D_z &=& D_X\sin{\beta}\cos{\alpha}+D_Y\sin{\beta}\sin{\alpha}+D_Z\cos{\beta}. \nonumber
\end{eqnarray}

\noindent Here angles $\alpha$ and $\beta$ define orientation of
the local coordinate system $(x,y,z)$ where Zeeman term in Eq.
(\ref{eqn:ham}) takes diagonal form $g\mu_{B}HS^z$ with
\begin{equation}
g=\sqrt{A^2+B^2+C^2}, \label{geff}
\end{equation}
where
\begin{eqnarray}
A&=&g_{XX}\sin\Theta\cos\phi+g_{XY}\cos\Theta+g_{XZ}\sin\Theta\sin\phi, \nonumber \\
B&=&g_{YX}\sin\Theta\cos\phi+g_{YY}\cos\Theta+g_{YZ}\sin\Theta\sin\phi, \nonumber \\
C&=&g_{ZX}\sin\Theta\cos\phi+g_{ZY}\cos\Theta+g_{ZZ}\sin\Theta\sin\phi \nonumber \\
\text{and} \nonumber \\
&\cos\alpha&=\frac{A}{\sqrt{A^2+B^2}}, \cos\beta=\frac{C}{\sqrt{A^2+B^2+C^2}}, \nonumber
\end{eqnarray}
here polar $\Theta$ and azimuthal $\phi$ angles define the direction
of external magnetic field, so that $\Theta$ and $\phi$ are counted
from $Y$ and $X$ axes, respectively, as during the $g$-tensor
recovery procedure.

Note that by  setting $J_{rung}=0$ and $\vect{D}_1=\vect{D}_2$ in
Eqn. (\ref{eqn:dH}) we immediately arrive to the known result for
1D Heisenberg chain with uniform DM interaction [see formula (16)
in Ref. \onlinecite{marsel}].

DIMPY has two inequivalent ladders with different $g$-tensors and
DM vectors. In accordance with crystal symmetry of the DIMPY (see
Sec. \ref{sect:sym}), the legs within same ladder are linked by
inversion, so that

\begin{equation}
\hat{\mathbf{g}}_{1}^{(k)}=\hat{\mathbf{g}}_{2}^{(k)},\;
\vect{D}_1^{(k)}=-\vect{D}_2^{(k)},\;(k=1\;\text{or}\;2),
\label{symrel1}
\end{equation}

\noindent here upper index ($k=1,2$) denotes the inequivalent
ladders and lower index enumerate legs od the ladder. The legs of
inequivalent ladders are linked by screw rotation along the second
order axis, hence
\begin{equation} \label{symrel2}
\begin{split}
\vect{D}_{l}^{(2)}=&C_{2}(Y)\vect{D}_{l}^{(1)},\\
{\hat{\mathbf{g}}}_{l}^{(2)}=C_2(Y){\hat{\mathbf{g}}}_{l}^{(1)}&C_2(Y)^{-1},\;(l=1,2),
\end{split}
\end{equation}

Orientation of the local axes is essentially different for inequivalent
spin ladders. Having known directions of the main axes of $g$-tensors (see Sec.
\ref{sec:g-tensor}), it's easy to find their components referred to
crystallographic axes
\begin{eqnarray}
\hat{g}^{(1,2)}=\begin{pmatrix} 2.128 & \mp0.12 & -0.008 \\
\mp0.12 & 2.214 & \pm0.005 \\
-0.008 & \pm0.005 & 2.038 \end{pmatrix},
\label{gten}
\end{eqnarray}

\noindent where upper(down) sign corresponds to the ladder with
upper(down) sign of $\vect{n}_g$.

Simulation of experimental data on the linewidth angular
dependence by Eq. (\ref{eqn:dH}) showed that the
Dzyaloshinskii-Moria interaction  describes the angular variation
of the linewidth in DIMPY well enough (within experimental error).
However model including Dzyaloshinskii-Moria interaction only
predicts value of linewidth which is systematically less then the
experimental values by about 12 Oe. This fact indicates there is
an additional (small compared to DM interaction) source of the
line broadening in DIMPY. The modelled values of the linewidth can
be reconciled with the experimental ones by adding isotopic
contribution $\Delta H_0=12$~Oe, which can be probably ascribed to
the residual spin-lattice relaxation, or by considering other
anisotropic spin-spin couplings. Contribution to the linewidth
from dipole-dipole interaction is quite small for DIMPY and at the
shortest distance between Cu ions ($r=a\approx7.5$~\AA{})
following conventional estimation\cite{Yamada1996} it does not
exceed $\sim~0.5$~Oe. Additional broadening in DIMPY can be
related to SAE interaction along legs and rungs of the spin
ladders which usually appear as further sources of ESR line
broadening beyond the dominant DM interaction.
\cite{marsel,Eremin2008} The contribution to the linewidth due to
SAE interaction is derived in Appendix \ref{app:SAE}.

Taking into account symmetry relations (Eqs. (\ref{symrel1}),
(\ref{symrel2}), (\ref{gten}), (\ref{symrel3})) and, for
definiteness setting $\vect{D}_{1}^{(1)}=\vect{D}$,
$\hat{\mathbf{A}}_1^{(1)}=\hat{\mathbf{A}}$, the fitting of
linewidth angular dependence yields $D_X=0.15$, $D_Y=-0.14$,
$D_Z=0.075$~K and almost diagonal exchange-tensor with components
$J_{XX}=0.08$, $J_{YY}=-0.03$, $J_{ZZ}=-0.05$, $J_{XY}=-0.015$~K
and $J_{XZ}=J_{YZ}=0$. During simulation the Lorentzian profile
with exponential wings was assumed, which implies $C=\pi\sqrt{2}$.

Directions of the found $\vect{D}$ vectors are shown on the Figure
\ref{fig:struct}. Components of the DM vector and SAE tensor given
above correspond to the ladder with the $g$-tensor main axis
orientation $\vect{n}_{g1}=(\sin\Theta\cos\phi;
\cos\Theta;\sin\Theta\sin\phi)\approx(-0.57; 0.82;0)$ (when
comparing with Fig.\ref{fig:struct} note that $\vect{n}_g$ and
$-\vect{n}_g$ are physically equivalent),  angle between
$\vect{D}$ and $\vect{n}_g$ vectors (reduced to
$(-\frac{\pi}{2};\frac{\pi}{2})$ range for convenience) is
approximately 23$^\circ$.  Magnitude of the obtained
Dzyaloshinskii-Moria vector $|\vect{D}|=0.22$~K agrees well with
the crude estimation above (section \ref{sec:width-t}).

As it is seen from Fig. \ref{fig:width-angular}, taking into
consideration only an exchange mechanisms of spin anisotropy
within the legs of ladders gives a good compliance with
experiment. However, it is necessary to stress out, that without
DM interaction SAE coupling only (see Appendix \ref{app:SAE})
totally failed to give a correct description of the angular
dependence of linewidth in DIMPY.

Our simulation shows that absolute value as well as angular
anisotropy of the linewidth are predominantly attributed by DM
interaction, while contribution to the linewidth due to SAE
interaction is relatively small compared to DM one and gives a weak
angular variation. Similar behavior of ESR linewidth with coexistent
contributions from DM and SAE interactions within $S=1/2$
antiferromagnetic chains was observed in hight symmetry crystal
structure KCuF$_3$ \cite{Eremin2008}.

The found  DM vector have not only transverse but also nonzero
\textit{longitudinal} (with respect to the Cu-Cu exchange bond)
component within the legs. Such result does't contradict with the
general rules for DM vector, established by Moriya
\cite{Moriya1960} based on general symmetry grounds for a pair of
exchange interacting ions. Moreover, a simple analysis of the
recovered $g$-tensors (see Sec. \ref{sec:g-tensor}) leads to the
same conclusion about direction of the DM vector. Since the axial
component of the $g$-tensor has a maximal value, then the ground
state orbitals of Cu$^{2+}$ ions (typically
"$\tilde{x}^2-\tilde{y}^2$"-like symmetry and $\tilde{z} ||
\vect{n}_{g}$) should predominantly lie within the plane
perpendicular to the main axis of a $g$-tensor, because the
maximal matrix element ($<\tilde{x}^2-\tilde{y}^2|
l_g|\tilde{x}\tilde{y}>=-2\imath$) relevant to spin-orbital
coupling appears only in the case when an external magnetic field
is applied parallel to the main axis of $g$-tensor. For the same
reason an effective DM vector predominantly should lie along the
main axis of the $g$-tensor. It should be noted that conventional
rule determining DM vector as
 $\vect{D}\propto \left[ n_{1} \times n_{2} \right]$,
 \cite{Keffer1962, Moskvin1977} where $n_1$ and
$n_2$ are unit vectors connecting a exchange interacting ions with
bridging ion, is not applicable in present case. Possible failure
of this rule was mentioned before in Ref. \onlinecite{Eremin1965},
referring to the features of exchange process through a two
bridging ions, which is also the case of DIMPY (see Fig.
\ref{fig:struct}).

Thus, analysis of  ESR linewidth allowed us to conclude that the
DIMPY is a rare case of compound in which DM vector has a component
along the line connecting the pair of exchange interacting ions.
This is  a consequence of  low crystal symmetry of DIMPY and
nontrivial orbital ordering.

\subsection{Low-temperature sub-components appearance.}

First, we recall main observations on the subcomponent appearance.
ESR components splits around 1~K into two sub-components, one of
which is much weaker. The splitting is best observed at
$\vect{H}||(X+Y)$ orientation. Position of the weaker
sub-component with respect to the stronger sub-component is
different for both ESR absorption components. Maximal splitting is
about 150~Oe and it decreases as the resonance field approaches
critical field, weaker subcomponent became unresolvable at the
fields above 2/3 of the critical field. Activation energies for
the stronger and weaker subcomponents are different.

All these observations can be explained  as an effect of the zero
field splitting of triplet sublevels. This effect was already
observed for various spin-gap magnets, e.g. TlCuCl$_3$,
\cite{glazkov-tlcucl3} or PHCC. \cite{glazkov-phcc} Anisotropic
interactions lift degeneracy of the $S=1$ triplet state and
frequencies of the  dipolar transitions $|S^z=+1 \rangle
\leftrightarrow |S^z=0 \rangle$ and $|S^z=-1 \rangle
\leftrightarrow |S^z=0 \rangle$ would become different. Here we
assume, which is perfectly valid for the case of DIMPY, that the
anisotropy is very small and spin projection on the field
direction $S^z$ can be considered as a good quantum number.
Therefore, in the presence of such an anisotropy the resonance
fields for  $|S^z=+1 \rangle \leftrightarrow |S^z=0 \rangle$ and
$|S^z=-1 \rangle \leftrightarrow |S^z=0 \rangle$ transitions in
the constant frequency ESR experiment would differ and ESR
absorption split into two sub-components.

Observed difference of the activation energies for the absorption
sub-components and dependence of the activation energy on the
microwave frequency used in the experiment is a direct consequence
of this explanation. The ESR intensity at low temperature is
determined by the population of the lowest sublevel. Hence, for
the $|S^z=+1 \rangle \leftrightarrow |S^z=0 \rangle$ transition
the activation energy is $\Delta \approx \Delta_0-g\mu_B
H_{res}=\Delta_0-h\nu$, being determined by the population of the
$|S^z=+1 \rangle$ sublevel (energy of this sublevel decrease with
field, see inset on Figure \ref{fig:int-final}). In the same time
the activation energy for the $|S^z=-1 \rangle \leftrightarrow
|S^z=0 \rangle$ remains constant (and equal to $\Delta_0$) since
the energy of $|S^z=0 \rangle$ sublevel is field independent. The
dependences of the activation energy on the microwave frequency of
the ESR experiment are described by this model parameter-free
using the zero-field gap value of 0.33~meV from the inelastic
neutron scattering experiment. \cite{dimpy2,dimpy-dave-prb}

Behaviour of the sublevels of the spin-gap magnet in the vicinity
of the critical field is a long-discussed problem. There is a
general macroscopic (or bosonic) approach of Refs.
\onlinecite{Affleck}, \onlinecite{farmar} and a 1D fermionic
approach of Tsvelik\cite{tsvelik} developed for the spin-chains.
Fermionic model of Tsvelik yields results formally equivalent to
the results of perturbation treatment of anisotropic
interactions.\cite{zaliznyak} Thus, within these approaches the
sublevels behave linearly in the vicinity of the critical field
and the splitting of the ESR subcomponent should be then field
independent. Bosonic model, on the contrary, predicts
square-root-like approach to the critical field for the low-energy
sublevel, while field dependence of the high-energy sublevel
remains linear in the vicinity of the critical field. Therefore,
sub-components splitting will change close to the critical field.
However, this nonlinearity of the bosonic model  extends only in
the small vicinity of the critical field $(H_c-H) \sim \Delta
E/\mu_B \sim \Delta H$, here $\Delta E$ is the zero field  triplet
sublevels splitting and $\Delta H \simeq 150$~Oe is the observed
sub-components splitting. We observe (Figure \ref{fig: max-split})
that the observed splitting is halved (compare 50.18~GHz and 26.30
~GHz curves at the Figure \ref{fig: max-split}) in the field of
about 2/3 of the critical field ($H_c \simeq 30$ kOe, zero-field
gap of 0.33~meV corresponds to the frequency of 80~GHz), i.e. well
below this nonlinearity range. This probably indicates that field
evolution of the split sub-components follows some other laws on
approaching the critical field. Similar behaviour of the ESR line
split by the uniform Dzyaloshinskii-Moria interaction was recently
reported for a quasi-1D antiferromagnet
Cs$_2$CuCl$_4$.\cite{timofey}

Under an assumption that the uniform Dzyaloshinskii-Moria
interaction along the legs of the ladder is responsible for the
observed splitting, anisotropy axis have to be aligned along the
$\vect{D}$ vector. We calculated effects of the DM coupling
perturbatively for the limiting case of strong-rung ladder (see
Appendix \ref{app:perturb}). Interdimer DM interaction mixes one
and two-particle excited $S^z=\pm 1$ states which results in the
triplet sublevels splitting by $\delta E=\frac{D^2}{2 J}$,
$S^z=\pm 1$ sublevels being shifted down. This corresponds to the
easy-axis anisotropy for the triplet excitations, $\vect{D}$
direction being the easy axis direction. Taking the magnitude of
the DM vector $D\approx 0.20$~K as estimated from the
high-temperature ESR linewidth analysis and substituting energy
gap of 0.33~meV as an exchange parameter of the perturbative model
we obtain an estimate of the sublevels splitting $\delta E \simeq
5$~mK which corresponds to sub-components splitting of about
40~Oe, factor of four less then the experimentally observed value.
However, perturbative treatment starting from the uninteracting
dimers is at best a qualitative model for a strong-leg ladder and
a detailed description of a strong-leg spin ladder with uniform
Dzyaloshinskii-Moria interaction needs a separate theoretical
effort.

We can not unambiguously determine type of the anisotropy from our
experimental observation since our setup does not allow to take an
angular dependence at He-3 temperature range. However, as it is
known from the formally similar problem of $S=1$ ion in a crystal
field \cite{Abragam,AltKoz} the effective anisotropy constant
changes monotonously with field rotating away from the anisotropy
axis $C_{eff}=\frac{C}{2}\left(3 \cos^2\xi-1\right)$, where $\xi$
is an angle counted from the anisotropy axis $z$ and anisotropy
$C$ enters spin Hamiltonian as $C ({S}^z)^2$. It is maximal at the
field parallel to the anisotropy axis, it change sign and
decreases by the factor of two at the orthogonal orientation of
the magnetic field and it turns to zero at a magic angle. Thus, as
splitting observed for the high-field component is larger then
that from the low-field component (approximately 150 Oe vs. 110
Oe, see Figure \ref{fig: max-split}) and weaker sub-component is
located on different side from the main sub-component, we find it
more likely that the high-field component corresponds to the
ladder with the magnetic field close to the true anisotropy axis.
In this case, as for the field applied close to anisotropy axis
the weaker subcomponent is located to the right from the stronger
subcomponent, the splitting of the triplet sublevels follows
easy-axis type of anisotropy energy of $S^z=\pm1$ states being
lower then energy of $S^z=0$ state in zero field.

However, this tentative identification of the anisotropy axis
deviates from the simple model of DM interaction only: as  the
vectors $\vect{D}$ and $\vect{n}_g$ are quite close for the given
ladder the low-field component (corresponding to the higher
longitudinal $g$-factor) should then  be closer to the anisotropy
axis. Possible reason for this deviation is the effect of
symmetric anisotropic exchange on the rungs of the ladder (see
Appendix \ref{app:perturb}). SAE coupling is smaller in magnitude,
but it enters to the triplet sublevels splitting linearly, why DM
contribution is quadratic. This is contrary to the linewidth
calculations where both couplings enter quadratically. Thus,
description of the subcomponents splitting probably lies beyond
the simple model with DM interaction only and requires accounting
for other anisotropic interactions.

\section{Conclusions.}
The strong-leg spin ladder system DIMPY is an established test
example of the Heisenberg spin ladder. However, anisotropic
spin-spin interactions, and in particular Dzyaloshinskii-Moria
interaction of intriguing geometry: uniform along the leg of the
ladder and exactly opposite on the other leg, give rise to a
family of interesting phenomena.

We have estimated parameters of Dzyaloshinskii-Moria interaction
from high-temperature data. We observe splitting of the ESR line at
low temperatures which is related to the zero-field splitting of the
triplet sublevels by the same interaction. Finally, we observe
series of  crossovers between different regimes of relaxation of
spin precession on cooling from room temperature to 400~mK.

We present qualitative explanations of our observations. Simple
geometry of the exchange couplings and anisotropic spin-spin
interactions makes DIMPY one of the few candidates for the
model-free microscopic description of the effects of anisotropic
interactions on the properties of a spin-gap magnet, which is
still awaiting for a theoretical effort.

\acknowledgements

We thank A.B.Drovosekov (Kapitza Institute) for the assistance with
ESR experiment above 77~K and  Prof.A.K.Vorobiev (M.Lomonosov Moscow
State University) for the possibility to perform  reference
measurements with X-band spectrometer. We thank K.Povarov, Prof.
M.V.Eremin and Prof. A.I.Smirnov for valuable and stimulating
discussions. Authors acknowledge usage of "Balls \& Sticks" software
to build crystal structure images.

The work was supported by Russian Foundation for Basic Research
Grant No.15-02-05918, Russian Presidential Grant for the Support
of the Leading Scientific Schools No.5517.2014.2. M .F.  work was
supported by the Russian Government Program of Competitive Growth
of Kazan Federal University. This work was partially supported by
the Swiss National Science Foundation, Division 2.

\appendix

\section{Linewidth contribution by a SAE
interaction along the legs.\label{app:SAE}}

Symmetric anisotropic exchange (SAE) coupling is allowed both on
the rungs and on the legs of the lader. Our aim here is to
demonstrate that SAE coupling can explain contribution of about
20\% of the total linewidth that can not be described by DM
coupling alone. We will focus here on a SAE coupling along the
legs of the ladder. We have checked that a SAE coupling along the
rung yields similar angular dependence and its contribution
differs only by some numerical scaling factor. However we expect
that contribution of the SAE couplings on the rungs should be
small in a strong-leg ladder since the overlapping of the orbitals
along the rung is smaller.

An expression for ESR linewidth due to a SAE coupling along the legs
of spin ladder is derived similarly as was done for a DM one in Sec.
\ref{sec:width-ang}, applying
\begin{equation}
{\cal H}_{SAE}=\sum_i \sum_{l=1,2}
\vect{S}_{l,i}\hat{\mathbf{A}}_l\vect{S}_{l,i+1}
\label{eqn:HanSAE}
\end{equation}
to the Eqs.(\ref{M2}), (\ref{M4}), (\ref{linewidth}) in the
framework of "method of moments", that in the high temperature limit
of $S=1/2$ leads to
\begin{widetext}
\begin{equation}
\Delta H_{\infty}^{SAE}=C\sum_{l=1,2}
\frac{[(2\lambda_{zz}-\lambda_{xx}-\lambda_{yy})^2+10(\lambda^2_{xz}+\lambda^2_{yz})
+(\lambda_{xx}-\lambda_{yy})^2+4\lambda^2_{xy}]_l}{8\sqrt{6}\mu_{B}\tilde{J}_Sg(\theta,\phi)},
\label{eqn:dHSAE}
\end{equation}
\end{widetext}
where $\tilde{J}_S=\sqrt{J_{leg}^2+2/3J_{rung}^2}$, and the
exchange-tensor components in the local coordinates,
$\lambda_{\eta\gamma}(\alpha,\beta)$, ($\eta,\gamma=x,y,z$) are
defined by transformation given in Ref. \onlinecite{eremins} [see
formulae A2-A4 therein] and, as usually, are expressed via the
intrinsic exchange parameters $J_{\mu\tau}$ of symmetric tensor
$\hat{\mathbf{A}}$ in crystallographic coordinates
($\mu,\tau=X,Y,Z$).

In accordance with crystal symmetry of the DIMPY (see Sec.
\ref{sect:sym}), symmetrical exchange tensors on the legs of same
ladder are equal, while on the legs of inequivalent ladders they are
related by a screw rotation along the second order axis, that is
\begin{equation} \label{symrel3}
\begin{split}
&\hat{\mathbf{A}}_{1}^{(k)}=\hat{\mathbf{A}}_{2}^{(k)},\;(k=1\;\text{or}\;2)\\
{\hat{\mathbf{A}}}_{l}^{(2)}=&C_2(Y){\hat{\mathbf{A}}}_{l}^{(1)}C_2(Y)^{-1},\;(l=1,2)
\end{split}
\end{equation}

Generally, taking into account relation of
$J_{\eta\gamma}=J_{\eta\gamma}$ for a symmetric tensor, an
exchange tensor has  six different components, constrained by $Tr
J_{\eta\gamma}=0$ condition. However, since the anisotropy of the
$g$-tensor and the tensor of SAE coupling originates from the same
spin-orbital interactions, accidental smallness of $g_{XZ}$ and
$g_{YZ}$ components allows to assume $J_{XZ}=J_{YZ}=0$ during the
fitting procedure. We have found, that the remaining four
components of the SAE tensor are enough to reproduce our data.

\section{Perturbative treatment of triplet sublevels splitting  by  uniform
DM and  SAE couplings.\label{app:perturb}}

Strong-rung $J_{rung}\gg J_{leg}$ limit allows to use
wavefunctions of the isolated dimers as a zero-order
approximation. For the dimer located at the $n$-th rung
wavefunction of the ground state is $\psi_{n0}=\frac{1}{\sqrt{2}}
(| \uparrow\downarrow \rangle_n-|\downarrow\uparrow\rangle_n)$ and
wavefunctions of the excited triplet are $\psi_{n11}=|
\uparrow\uparrow\rangle_n$, $\psi_{n10}=\frac{1}{\sqrt{2}} (|
\uparrow\downarrow \rangle_n+|\downarrow\uparrow\rangle_n)$ and
$\psi_{n1-1}=| \downarrow\downarrow\rangle_n$. Wavefunction of the
collective ground state is $\Psi^{(0)}=\prod_p \psi^{(p)}_0$ and
single-particle excited states with the excitation at the n-th
dimer can be build as
\begin{equation}\label{eqn:wavefunctions}
\Psi^{(1)}_{n11}=\prod _{p=0}^{n-1} \psi_{p0} \times
\psi_{n11}\times \prod _{p=n+1}^{N}\psi_{p0}
\end{equation}
\noindent and similarly for other spin projections. Manyparticle
excited state can be constructed similarly keeping in mind
hard-core repulsion as only one excited state per dimer is
allowed.

One-particle states are $N$-fold degenerated, this degeneration
will be lifted by interdimer exchange coupling $J_{leg}$ giving
rise to excitations dispersion.

We consider effect of interdimer DM interaction (\ref{eqn:Hanis})
with DM vectors oppositely aligned on the legs of the ladder. This
configuration conserves symmetry axis (direction of the DM vector,
which we will use as $z$-direction), thus excitations will have
well defined $S^z$ values. Interdimer DM interaction can then be
expressed as

\begin{eqnarray}
  V&=&\frac{D}{2\imath}\sum_n[S_{1,n}^{-}{\hat
  S}_{1,n+1}^{+}-S_{1,n}^{+}S_{1,n+1}^{-}-\nonumber\\
  &&-S_{2,n}^{-}{\hat
  S}_{2,n+1}^{+}+S_{2,n}^{+}S_{2,n+1}^{-}]\label{eqn:HDM2}
\end{eqnarray}

By applying this operator to the ground state and to one-particle
excited states we obtain
\begin{eqnarray}
  V \Psi^{(0)}&=&0 \label{eqn:V-gs}\\
  V \Psi^{(1)}_{n10}&=&0 \label{eqn:V-10}\\
  V \Psi^{(1)}_{n11}&=&\frac{D}{2\imath}(-\Psi^{(2)}_{n10;(n+1)11}+\Psi^{(2)}_{n10;(n-1)11}) \label{eqn:V-11}
\end{eqnarray}

\noindent here $\Psi^{(2)}_{n10;m11}$ are two-particle excited
states with $S=1, S^z=0$ excitation on the $n$-th rung and
$S=1,S^z=1$ excitation on the $m$-th rung.

Thus, interdimer DM interaction mixes $S^z=\pm 1$ single particle
excited states with $S^z=\pm 1$ two particle excited states. This
mixing results in the second-order perturbative correction to the
energy of the single-particle state $\delta
E=-\frac{D^2}{2J_{rung}}$, $S^z=\pm 1$ being shifted down. As this
correction is the same for all $S^z=\pm 1$ states, weak interdimer
Heisenberg coupling $J_{leg}$ will not affect it.

This result differs from the effect of intradimer DM interaction
which mixes $S=1, S^z=0$ state with $S=0$ state and shifts the
energy of the $S=1, S^z=0$ up by $\frac{D^2} {4 J_{rung} }$.
However both interdimer and intradimer DM interaction results in
the easy-axis anisotropy for the triplet excitations (energy of
the $S^z=\pm 1$ states is lower then the energy of the $S^z=0$
state).

Note also, that the effective anisotropy for the triplet
excitations is easy axis, while usually anisotropy due to DM
interaction (e.g., anisotropy of the order parameter in the
ordered state of an antiferromagnet) is of easy plane type. This
"inversion" of anisotropy seems to be a common feature of all
spin-gap magnets: it was obtained by perturbative analysis of the
role of single-ion anisotropy in Haldane magnet \cite{Gollineli}
and was observed in a Haldane magnet
PbNi$_2$V$_2$O$_8$,\cite{pbnio} similar inversion of the
anisotropy type between the anisotropy of triplet excitations and
order parameter anisotropy in a field-induced ordered phase of a
spin-gap magnet follows from macroscopic approach\cite{farmar}.

Effect of the symmetric anisotropic coupling on a strong-rung
ladder can be considered similarly. Our aim here is to illustrate
that its contribution is linear on coupling parameter along the
rung so we consider simple axial SAE coupling in the form

\begin{equation}\label{eqn:sae-perturb}
 {V}=A_{leg}\sum_{n,j}
 S^z_{j,n}S^z_{j,n+1}+A_{rung}\sum_n
 S^z_{1,n}S^z_{2,n}
\end{equation}

\noindent here $n$ enumerates rungs (dimers) and $j$ enumerates
legs of the ladder (spins in the dimer), $A_{leg}$ and $A_{rung}$
are SAE coupling constants along the leg and rung of the ladder.

By applying this operator to the ground state and to the
one-particle excited states one can ascertain that SAE coupling
along the legs mixes ground sate and single-particle states with
two-particle states and will give some corrections in the second
order of perturbations, while SAE coupling on the rungs gives
energy corrections already in the first order on the coupling
parameter: energies of the ground state and of the $S^z=0$
component of triplet state are shifted (per dimer) by
$-\frac{1}{4}A_{rung}$, while energies of the $S^z=\pm1$
components are shifted by $+\frac{1}{4}A_{rung}$. Zero field
splitting of triplet levels appears, its type (easy axis or easy
plane) depends on the coupling parameter $A_{rung}$ sign, which
can be both positive and negative.


\begin{thebibliography}{10}

\bibitem{kolezhukmikeska} H.J.Mikeska and A.K.Kolezhuk, Lect. Notes Phys., \textbf{645}, 1-83 (2004)

\bibitem{ntenp} M. Hagiwara, L. P. Regnault, A. Zheludev, A. Stunault, N. Metoki,
T. Suzuki, S. Suga, K. Kakurai, Y. Koike, P. Vorderwisch, and
J.-H. Chung, Phys. Rev. Lett. 94, 177202 (2005)

\bibitem{Affleck} Ian Affleck, Physical Review B \textbf{46},
9002 (1992)

\bibitem{tsvelik} A.M. Tsvelik, Physical Review  B \textbf{42}, 10499 (1990)

\bibitem{farmar}A.M. Farutin and V.I. Marchenko, Zh.Eksp.Teor.Fiz. \textbf{131} 860
(2007) (JETP \textbf{104} 751 (2007))

\bibitem{Kolezhuk} A.K. Kolezhuk, V.N. Glazkov, H. Tanaka, and A. Oosawa
Physical Review  B \textbf{70}, 020403 (2004)


\bibitem{povarov} K. Yu. Povarov, A. I. Smirnov, O. A. Starykh, S.
V. Petrov, and A. Ya. Shapiro, Physical Review Letters
\textbf{107}, 037204 (2011)


\bibitem{dimpy1}A. Shapiro, C. P. Landee, M. M. Turnbull, J. Jornet, M. Deumal,
J. J. Novoa, M. A. Robb, and W. Lewis, J. Am. Chem. Soc.
\textbf{129}, 952 (2007)

\bibitem{dimpy2} T. Hong, Y. H. Kim, C. Hotta, Y. Takano, G. Tremelling, M. M.
Turnbull, C. P. Landee, H.-J. Kang, N. B. Christensen, K. Lefmann,
K. P. Schmidt, G. S. Uhrig, and C. Broholm, Phys. Rev. Lett.
\textbf{105}, 137207 (2010).

\bibitem{dimpy-dave-prb} D. Schmidiger, S. M\"{u}hlbauer, S. N. Gvasaliya, T. Yankova, and A. Zheludev
Phys. Rev. B \textbf{84}, 144421 (2011)


\bibitem{dimpy-dave-prl}D. Schmidiger, P. Bouillot, S. M\"{u}hlbauer, S. Gvasaliya, C. Kollath, T. Giamarchi, and A. Zheludev
Phys. Rev. Lett. \textbf{108}, 167201 (2012)

\bibitem{dimpy-magn} J. L. White, C. Lee, \"{O}. G\"{u}naydin-\c{S}en, L. C. Tung, H. M. Christen, Y. J. Wang, M. M. Turnbull, C. P. Landee,
R. D. McDonald, S. A. Crooker, J. Singleton, M.-H. Whangbo, and J.
L. Musfeldt Phys. Rev. B \textbf{81}, 052407 (2010)

\bibitem{eremins} R. M. Eremina, M. V. Eremin, V. N. Glazkov, H.-A. Krug von Nidda and A.
Loidl, Physical Review B \textbf{68}, 014417 (2003)

\bibitem{dimpy-highfieldESR} Hitoshi Ohta, Tatsuya Yamasaki, Susumu Okubo, Takahiro Sakurai,
Masashi Fujisawa, Hikomitsu Kikuchi, Journal of Physics:
Conference Series \textbf{320} 012026  (2011)

\bibitem{anderson-stat}
{P.W.Anderson}, Journal of the Physical Society of Japan
\textbf{9} (1954), 316

\bibitem{exnar2}
{L.H.Piette}, {W.A.Anderson}, Journal of Chemical Physics
  \textbf{34} (1959), 899

\bibitem{marsel} M. A. Fayzullin, R. M. Eremina, M. V. Eremin, A. Dittl, N. van Well, F. Ritter, W. Assmus, J. Deisenhofer, H.-A. Krug von Nidda, and A. Loidl
Physical Review B \textbf{88}, 174421 (2013)

\bibitem{monika} M. Heinrich, H.-A. Krug von Nidda, A. Krimmel, A. Loidl, R. M.
Eremina, A. D. Ineev, B. I. Kochelaev, A. V. Prokofiev, and W.
Assmus, Physical Review B \textbf{67}, 224418 (2003).

\bibitem{Abragam} A. Abragam, B. Bleaney, Electron paramagnetic resonance of transition ions, Clarendon Press, Oxford (1970)

\bibitem{AltKoz}  S.Al'tshuler, B.Kozyrev,  Electron Paramagnetic
Resonance, Academic Press (1964)

\bibitem{Sato} S.C.Furuya and M.Sato, Journ. Phys. Soc. Japan
\textbf{84}, 033704 (2015)

\bibitem{oshikawa} M. Oshikawa and I. Affleck, Physical Review B \textbf{65},
134410(2002)


\bibitem{glazkov-tlcucl3}
V.N.Glazkov, A.I. Smirnov, H. Tanaka and A. Oosawa, Physical
Review B, \textbf{69} 184410 (2004).

\bibitem{glazkov-phcc} V. N. Glazkov, T. S. Yankova, J.
Sichelschmidt, D. H\"{u}vonen, A. Zheludev,  Physical Review B,
\textbf{85}, 054415 (2012)

\bibitem{chestnut} D.B.Chestnut and W.~D.~Phillips.
Journal of Chemical Physics, \textbf{35}, 1002 (1961).



\bibitem{Anderson1953} P. W. Anderson and P. R. Weiss,
Rev. Mod. Phys., \textbf{25}, 269 (1953).

\bibitem{Castner1971} T. G. Castner, Jr. and Mohidar S. Seehra,
Phys. Rev. B, \textbf{4}, 38 (1971).

\bibitem{VanVleck1948} J. H. Van Vleck,
Phys. Rev., \textbf{74}, 1168 (1948).


\bibitem{Zakharov2008}
D.V. Zakharov et al., in Quantum Magnetism, edited by
B. Barbara et al. (Springer, Dordrecht, 2008).

\bibitem{Huber1999} D. L. Huber, G. Alejandro, A. Caneiro, M. T. Causa, F. Prado, M. Tovar
Phys. Rev. B, \textbf{60}, 12155 (1999)

\bibitem{Yamada1996}
I. Yamada, M. Nishi, J. Akimutsu J. Phys.: Condens. Matter,
\textbf{8}, 2625 (1996).

\bibitem{Eremin2008}
M.V. Eremin, D.V. Zakharov, H.-A. Krug von Nidda, R.M. Eremina, A.
Shuvaev, A. Pimenov, P. Ghigna, J. Deisenhofer and A. Loidl
Physical Review Letters, \textbf{101}, 147601 (2008).

\bibitem{Moriya1960} T. Moriya,
Phys. Rev., \textbf{120}, 91 (1960).


\bibitem{Moskvin1977} A. S. Moskvin and I. G. Bostrem,
Sov. Phys. Solid State, \textbf{19}, 1532 (1977).

\bibitem{Keffer1962} F. Keffer,
Phys. Rev., \textbf{126}, 896 (1962).


\bibitem{Eremin1965}
M. V. Eremin, in Spektroskopiya Kristallov, edited by
A. A. Kapliyanski (Nauka, Leningrad, 1965).



\bibitem{zaliznyak}L.-P. Regnault, I.A. Zaliznyak and S.V. Meshkov
J.Phys.:Condens.Matter \textbf{5} L677 (1993)

\bibitem{timofey} A. I. Smirnov, T. A. Soldatov, K. Yu. Povarov, A. Ya.
Shapiro, Phys. Rev. B \textbf{91}, 174412 (2015)

\bibitem{Gollineli} O. Golinelli, Th. Jolicoeur, and R. Lacaze, J. Phys.: Condens.
Matter \textbf{5}, 7847 (1993)

\bibitem{pbnio} A. I. Smirnov, V. N. Glazkov, T. Kashiwagi, S. Kimura, M. Hagiwara, K. Kindo, A. Ya. Shapiro, and L. N.
Demianets, Physical Review B, \textbf{77}, 100401(R) (2008)
\end{thebibliography}
\end{document}